\documentstyle[twocolumn,prl,aps,epsfig]{revtex}
\newcommand{\beq}{\begin{eqnarray}}
\newcommand{\eeq}{\end{eqnarray}}

\begin{document}
\draft

\title
{Electron Quasiparticles Drive the Superconductor-to-Insulator Transition
in Homogeneously Disordered Thin Films}
\author{Philip Phillips and Denis Dalidovich}
\vspace{.05in}

%
\address
{Loomis Laboratory of Physics\\
University of Illinois at Urbana-Champaign\\
1100 W.Green St., Urbana, IL, 61801-3080}

%

\address{\mbox{ }}
\address{\parbox{14.5cm}{\rm \mbox{ }\mbox{ }
Transport data on Bi, MoGe, and PbBi/Ge homogeneously-disordered thin
films
demonstrate that the critical resistivity, $R_c$, at the nominal
insulator-superconductor transition is linearly proportional to the
normal sheet resistance, $R_N$.  In
addition, the critical magnetic field scales linearly with the
superconducting energy
gap and
is well-approximated by $H_{c2}$.  Because $R_N$ is determined at high
temperatures and $H_{c2}$ is the pair-breaking field, the two immediate
consequences are: 1) electron-quasiparticles populate the insulating side
of the transition and 2) standard phase-only models
are incapable of describing the destruction of the superconducting state.
As gapless electronic excitations populate the insulating state, the
universality
class is no longer the 3D XY model, thereby relaxing the constraints
that this model imposes on the critical exponents
 as well as on the critical resistance, namely $R_c\equiv R_Q=h/4e^2$.
The lack of a unique critical resistance
in homogeneously disordered films can be understood in this context.   In
light of the recent experiments
which observe an intervening metallic state separating the insulator
from the superconductor in homogeneously disordered
MoGe thin films, we argue that the two transitions that accompany the
destruction of superconductivity are 1)
superconductor to Bose metal in which phase coherence is lost and 2) Bose
metal
to localized electron insulator via pair-breaking.}}
\address{\mbox{ }}
\address{\mbox{ }}

\maketitle

\begin{figure}
\begin{center}
\epsfig{file=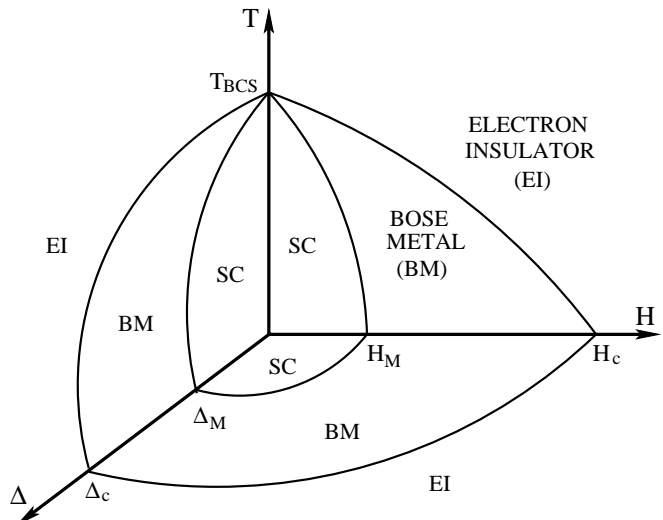, height=7.0cm}
\caption{Global phase diagram for destruction of superconductivity 
in disordered thin films.  At $\Delta_M$ and $H_M$,
phase coherence is lost giving rise to a Bose metal phase in which
the Cooper pair amplitude remains intact.  The pair amplitude vanishes
at the upper field $H_c$ and the disorder $\Delta_c$ yielding a
localized electron insulator.}
\label{pdiag}
\end{center}
\end{figure}

In an issue honouring the work of Michael Pollak,
certain topics such as the Coulomb gap and
the interplay between disorder and long-range Coulomb interactions
naturally come to mind.   Nonetheless, we would
like to focus on the insulator-superconductor transition
in homogeneously disordered thin films. The impetus for this work is the
recent
observation (Mason and Kapitulnik 1999,
Mason and Kapitulnik 2000)
in homogeneously disordered thin films of MoGe
that two phase
transitions are encountered in the magnetic field-tuned
transformation of the superconducting phase into an insulator.
We argue
here that the two transitions correspond to superconductor to Bose metal
in which Cooper pairs lack phase coherence and Bose metal to
localized electron
insulator as illustrated in the global phase
diagram in Fig. (\ref{pdiag}).   The latter interpretation is consistent with the extensive
work of Dynes, Valles and co-workers,
(Dynes, Garno and Rowell 1978, Dynes {\it et al.} 1994,
Barber {\it et al.} 1994, Valles, Dynes and Garno 1992,
Merchant {\it et al.} 2001, Hsu, Chervenak and Valles 1995,
Chervenak and Valles 2000, Chervenak and Valles 1999,
Valles {\it et al.} 2000, Hsu 1995, Chervenak 1998, Kouh 1999)
who have argued tirelessly
that the insulator-superconductor transition
in homogeneously disordered films is driven by amplitude fluctuations or
equivalently pair-breaking.  While it comes as no surprise
that Cooper pairs break apart for sufficiently strong disorder or magnetic fields, it is not widely accepted that this
state of affairs obtains generically at the nominal
superconductor-insulator transition in homogeneously disordered thin films.
That the transport and tunneling data lead necessarily to this conclusion and that loss of
phase coherence leads to a Bose metal phase (rather than a Cooper pair insulator) are the two key results
of this work. 
As Michael Pollak's work has been motivated primarily by experiments
on disordered 2D systems, insulator-superconductor transitions (IST)
in disordered thin films certainly intersect his work.

From the earliest experiments
(Dynes, Garno and Rowell 1978, Dynes {\it et al.} 1994,
Barber {\it et al.} 1994, Valles, Dynes and Garno 1992,
Merchant {\it et al.} 2001, Hsu, Chervenak and Valles 1995,
Chervenak and Valles 2000, Chervenak and Valles 1999,
Valles {\it et al.} 2000, Hsu 1995, Hebard and Paalanen 1990,
Liu and Goldman 1994, Jaeger {\it et al.} 1989, Yazdani and Kapitulnik 1995,
Goldman and Markovic 1998)
on superconducting thin metal alloy films, a simple picture emerged
for the magnetic-field or disorder-tuned transition to the insulating state.
Namely, the transition
is continuous with a well-defined critical field, $H_c$, below which
superconductivity occurs and above which an insulator obtains.
At the critical field (or the critical film thickness), the resistivity
is independent of temperature.  
\begin{figure}
\begin{center}
\epsfig{file=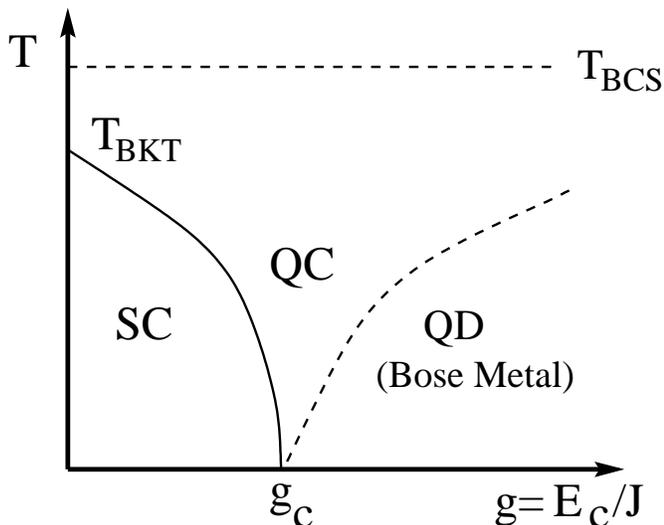, height=7.0cm}
\caption{Phase diagram for an array of Josephson
junctions as a function of temperature
and the quantum fluctuations, $g$.   For
 Josephson-junction arrays, $g$ is given
by the ratio of the charging energy, $E_c$, to the Josephson coupling,
$J$.
$T_{\rm BKT}$ is the Berezinskii- Kosterlitz-Thouless temperature at which phase
coherence
obtains.  $g_c$ defines the critical
value of the phase disorder to destroy
the superconducting phase.  QC refers to the quantum critical
regime in which the inverse correlation length is linear in the
temperature.
In the quantum disordered regime, QD, a gap appears to all excitations.
Phase
coherence is absent in this regime. Although this phase is gapped,
the dc conductivity remains finite as shown by Dalidovich
and Phillips (2000).
$T_{\rm BCS}$ defines the mean-field
temperature at which the pair amplitude first becomes non-zero.}
\label{vortex}
\end{center}
\end{figure}
\noindent While phase or pair amplitude fluctuations
both destroy superconductivity, the dominant model that was used to
explain all of the original experiments on the IST was the
phase-only 3D XY model or equivalently the charging model for
a Josephson junction array (JJA) (Fisher 1990,
Fisher, Grinstein and Girvin 1990, Schakel 2000).
Superconducting islands with charging energy $E_C$
linked together by a Josephson pair current, $J$, constitute the JJA
model. Such a model is expected to apply to any system, granular or otherwise,
in which Cooper pair formation and phase coherence occur at distinct
energy scales.
As illustrated in Fig. (\ref{vortex}),
such a model possesses a zero-temperature quantum critical point
driven solely by fluctuations of the quantum mechanical phase of
the order parameter.  On the superconducting side of the transition,
Cooper pairs form a phase-coherent state in which the resistivity
vanishes. On the insulating side, phase coherence is lost and all
excitations become gapped. The energy scale for the gap is set by the
inverse correlation length. The excitations are purely
bosonic, however, as the pair-amplitude is assumed to remain frozen
below some mean-field temperature, $T_{\rm BCS}$.
Consequently, electron-like quasiparticles arising
from pair amplitude fluctuations are strictly
absent from phase-only models.
Close to the finite $T_c$ line in Fig. (\ref{vortex}),
the elementary excitations correspond to vortex-antivortex pairs.  Such
excitations remain bound below $T_c$ but unbind above.  As a result of
the inherent duality between Cooper pairs and vortices, the
superconducting state can be thought of as a Cooper pair condensate
with localized vortices whereas in the quantum-disordered regime,
vortices condense but Cooper pairs remain
localized.

In a perpendicular magnetic field, vortices of only one vorticity are
possible in the superconducting phase.  Such vortices interact
logarithmically resulting in the stabilization
of an Abrikosov vortex lattice.  Finite randomness, however,
disrupts the lattice leading to glassy ordering. Thus, at sufficiently
weak magnetic fields, vortices are localized by
both logarithmic Coulomb repulsions and disorder, implying that
superconductivity is preserved at $T=0$.  At low temperatures and
non-zero magnetic fields, an exponentially
small resistivity is expected on the superconducting side as a result
of thermally assisted quantum tunneling of vortices.
In fact, it is the logarithmic Coulomb interaction between vortices in
the presence of disorder that conspires to
yield the Mott variable-range hopping form for the resistivity
$R(T) \propto \exp{-C|\ln T|/T}$ (Efros 1976).  Experimentally,
however, a leveling (Mason and Kapitulnik 1999,
Mason and Kapitulnik 2000, Ephron {\it et al.} 1996,
Geerligs {\it et al.} 1989) of the resistivity rather than an exponential
attenuation has been observed in finite field.
The saturation value of the resistivity is strongly dependent on
the magnetic field and implies
that a metallic phase intervenes between the superconductor and the
eventual insulator at $H_c$.  More recent
experiments (Mason and Kapitulnik 2000) have shown that
below a lower value of the magnetic field, $H_M$, the resistivity does
in fact vanish.
For the MoGe films, $H_M=.18T$ whereas $H_c\approx 2.5T$ indicating
that the two transitions are driven by fundamentally different physics.
The emergence of an intervening metallic phase is not restricted
only to the magnetic field-tuned problem.  In fact, in the disorder-tuned
transition, the resistivity has also been observed
(Jaeger {\it et al.} 1989)
to level at low temperatures for weak to intermediate values
of the disorder.  The saturation
value of the resistivity ranges from $.001k\Omega$ to $100k\Omega$.
Hence, the metallic phase is robust.

Within the standard dirty boson picture in which
bosons exist on both sides of the transition, two permissible
values for the resistivity are anticipated:
1) zero (superconductor) or 2) infinity (insulator).
However, we have shown (Dalidovich and Phillips 2000) recently
that the anticipated insulating state
is actually a metal (see Fig. (\ref{vortex})) with a universal dc
conductivity given by $(2/\pi)4e^2/h$.  While this result is rooted in
the noncommutativity (Damle and Sachdev 1997) of the temperature and
frequency tending to zero limits
of the conductivity, it has a simple physical interpretation.
In the quantum-disordered regime, the population of bosons
is exponentially suppressed; however, so is the scattering
rate between bosons.  But because the conductivity
is a product of the density and the scattering time,
the exponentials cancel, giving rise to a finite
conductivity at $T=0$.  Hence, the observation of a metallic state
is not entirely inconsistent with phase-only models as depicted in
Fig. (\ref{vortex}).  In fact, our work establishes that the Bose metal
is a generic
feature of phase-only models.
However, the occurrence of two phase transitions in MoGe
(Mason and Kapitulnik 1999, Mason and Kapitulnik 2000)
is inconsistent with
the phase-only model which possesses only a single critical point
signaling the termination of phase coherence.  Clearly,
both transitions cannot be caused by loss of phase coherence. 
In addition, experimentally, resistivities at criticality ranging
from $0.1R_Q$ to $3R_Q$ have been observed in
homogeneously disordered films (Liu and Goldman 1994),
with $R_Q=h/4e^2$. At criticality, Cooper pairs on the brink of forming
a phase coherent state are expected
(Fisher 1990, Fisher, Grinstein and Girvin 1990)
to diffuse with the quantum of conductance $\sigma_Q=4e^2/h$ for
charge $2e$ bosons.  Hence, such deviations
from $\sigma_Q$ and the observation of two transitions suggest that
physics beyond phase-only models might be relevant for homogeneously
disordered films.

Based on an analysis of the transport experiments
(Barber {\it et al.} 1994, Valles, Dynes and Garno 1992,
Hsu, Chervenak and Valles 1995, Yazdani and Kapitulnik 1995,
Goldman and Markovic 1998)
we suggest that the transition occurring at $H_c$ in MoGe as well
as the traditionally-studied IST in homogeneously disordered thin flims
involves an insulator populated with electron-like
quasiparticles as originally argued by Dynes,
Valles, and co-workers
(Dynes, Garno and Rowell 1978, Dynes {\it et al.} 1994,
Barber {\it et al.} 1994, Valles, Dynes and Garno 1992,
Merchant {\it et al.} 2001, Hsu, Chervenak and Valles 1995,
Chervenak and Valles 2000, Chervenak and Valles 1999,
Valles {\it et al.} 2000, Hsu 1995). A transition
driven by Cooper pair breaking lies outside the 3D XY universality
class and hence is not constrained to exhibit the unique value
of $h/4e^2$ at criticality.
For granular films, 
phase-only models appear to be adequate primarily because 
 $R_c\approx h/4e^2$ at criticality and the
single-particle energy gap (Barber {\it et al.} 1994)
 remains finite and unattenuated from the bulk
superconducting value on the insulating side of the transition.  We
propose that measurements of the superfluid density at both transitions
observed recently in MoGe will ultimately
determine which is of the Berezinskii-Kosterlitz-Thouless
(Berezinskii 1971, Kosterlitz and Thouless 1973) form.

\begin{figure}
\begin{center}
\epsfig{file=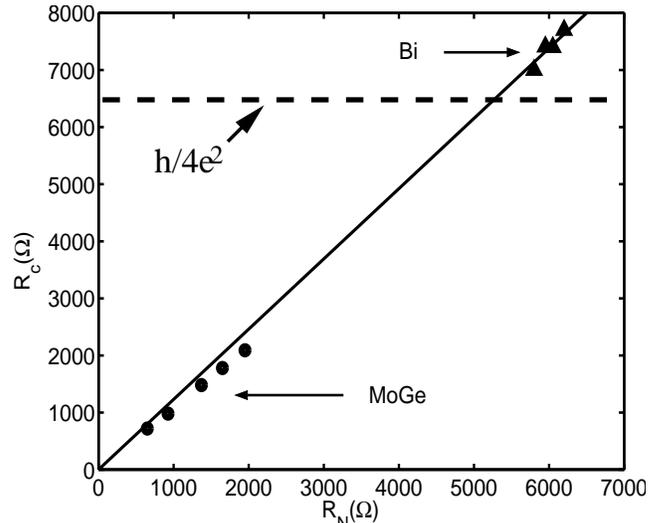, height=7.0cm}
\caption{Critical resistance, $R_c$ as a function of the normal
state resistance, $R_N$ for Bi (Goldman and Markovic 1998)
and MoGe (Yazdani and Kapitulnik 1995)
samples.  In both Bi and MoGe, an applied magnetic field
tunes the transition.  The lack of a universal value at the transition
and the correlation of $R_c$ with $R_N$ is an indication that the
transition
to the insulating state has electron-like quasiparticles, indicative of
the high temperature normal state. The dashed
line indicates the predicted value of the critical resistance for charge
2e bosons of the phase only model.  This value is 
seen in granular films but clearly
not in homogeneously disordered films including
$InO_x$ (Hebard and Paalanen 1990).}
\label{rnrc}
\end{center}
\end{figure}
To support our central claim that the time-honored IST
in homogeneously disordered films is accompanied by a proliferation of
electron-like quasiparticles, we compare the value of the resisitivity
at criticality with the resistivity in the normal state of the films.
Shown in Fig.
(\ref{rnrc}) is such a comparison for Bi and MoGe thin films.
The normal state resistance, $R_N$, was extracted
(Goldman and Markovic 1998, Yazdani and Kapitulnik 1995)
from transport measurements at $T>T_{\rm BCS}$.  Hence, only electron-like
excitations
abound in this temperature range. The critical resistance, $R_c$, is
extracted much below $T_{\rm BCS}$ and represents the limiting
value of the resistance at the single crossing point of $R(T)$
vs tuning parameter
as $T\rightarrow 0$.  In both the Bi and MoGe thin films, an
applied magnetic field induces
the IST. As is evident from the data shown,
for both Bi and MoGe, $R_c\propto R_N$. Unless a
transition to the normal ground state occurs at $R_c$, there is no
reason to expect that $R_c\propto R_N$.
We conclude then that because $R_N$ is associated with high temperature
physics in which no Cooper pairs exist, the fact that $R_c\propto R_N$
signifies that
the insulating state must have electron-like quasiparticles in
homogeneously disdordered thin films.
Further, in all cases, $R_c$ slightly exceeds
$R_N$ and ranges between $900\Omega$ and $8000\Omega$. 
In contrast,
granular films (Hebard and Paalanen 1990) display a critical resistance that is in accordance
with the prediction of phase-only models as the dashed line
in Fig. (\ref{rnrc}) attests.  Consequently, the non-universal value of $R_c$ 
in homogeneously disordered films implies that physics beyond phase-only
models must be at work.

\begin{figure}
\begin{center}
\epsfig{file=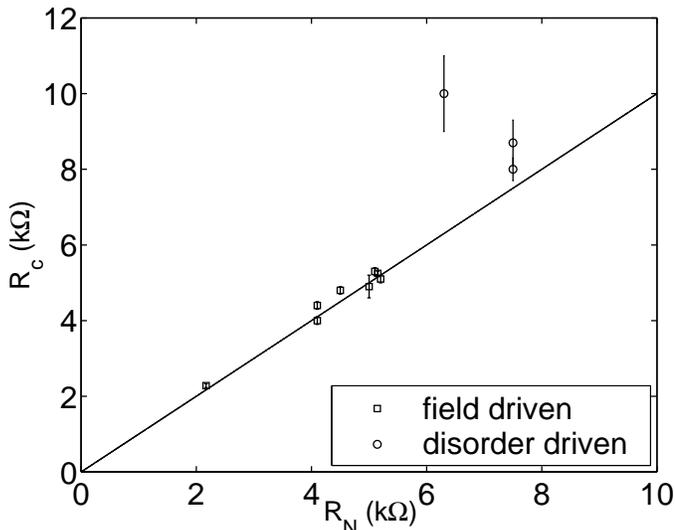, height=7.0cm}
\caption{Critical resistance, $R_c$ as a function of the normal
state resistance, $R_N$ for Bi/Sb (disorder-tuned)
(Chervenak and Valles 1999, Chervenak 1998, Kouh 1999)
and PbBi/Ge and Bi/Sb (field-tuned)
(Hsu, Chervenak and Valles 1995, Chervenak 1998, Hsu 1995)
samples. The linear relationship between $R_c$ and $R_N$ is consistent with
the Bi and MoGe data in Fig. (2).  This suggests that the IST
in homogeneous films is driven the high temperature physics of the normal
state.}
\label{vallesdata}
\end{center}
\end{figure}
\begin{figure}
\begin{center}
\epsfig{file=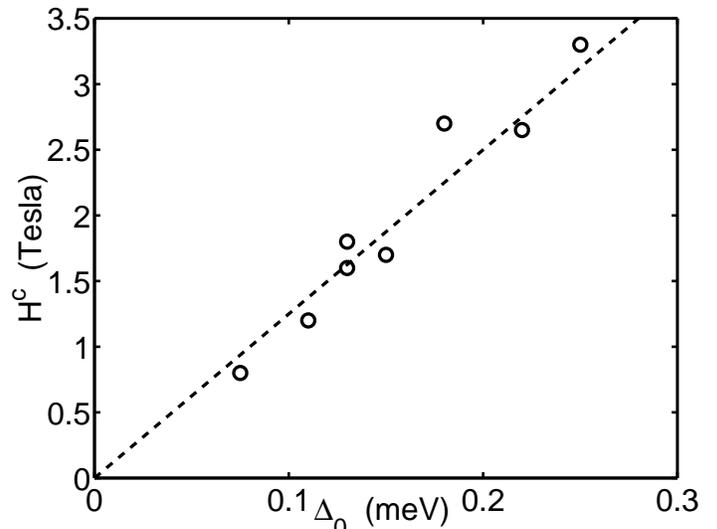, height=7.0cm}
\caption{Critical magnetic field $H_c$ versus the energy
gap, $\Delta_0$ for PbBi/Ge films.  Redrawn from S. -Y. Hsu's
thesis (Hsu 1995)}
\label{tunn}
\end{center}
\end{figure}

Further corroborating data on the homogeneously disordered
Bi/Sb and PbBi/Ge films are shown in Fig. (\ref{vallesdata}).
The error bars represent the inherent error in extracting $R_c$.
For the disorder-tuned transition,
the error bars are naturally larger than in the field-tuned case as a
result of the uncertainties in measuring the thickness of the sample.
For the field-tuned transition, the correlation between $R_c$ and
$R_N$ is striking and systematic.
Further, $R_c$ is not well approximated by $R_Q$.  These facts
are highly suggestive that the insulating state in
PbBi/Ge contains unpaired electrons as in the normal state
of the films.  However, the data which clinch this scenario
are the tunneling measurements which probe the single particle energy
gap. The emergence of electronic excitations
on the insulating side would imply that the Cooper pair amplitude
should vanish at the IST.
Within such a scenario, the critical
magnetic field is expected to be equal to $H_{c2}$ and as
a consequence should scale linearly with the gap in the single particle
spectrum.  Data (Hsu 1995) showing clearly that the gap in the
single particle spectrum inferred from tunneling scales linearly with the
critical field, $H_c$, are shown in Fig. (\ref{tunn}).  Such a
correspondence is expected within BCS theory
(Bardeen, Cooper and Schrieffer 1957),
provided that $H_c$ corresponds to $H_{c2}$.  Hence, the linear
relationship found here provides direct evidence that amplitude
fluctuations drive the transition to the insulating state as has been noted
previously by Valles and his co-workers
(Hsu, Chervenak and Valles 1995, Valles {\it et al.} 2000, Hsu 1995).
In fact, Valles and co-workers (Hsu, Chervenak and Valles 1995)
have determined that at $H_c$, the density of states at the Fermi energy
is 80$\%$ of its value in the normal state. Consequently, a large fraction
of the sample contains
no Cooper pairs.  This observation is also consistent
with the series of experiments by Dynes and co-workers
(Valles, Dynes and Garno 1992), in which the energy
gap has been observed to vanish at the nominal critical field, $H_c$.
While the average gap might vanish at the transition, pairs can still
exist as a result of fluctuations of the pair amplitude.  Consequently,
above $T_c$, Aslamazov-Larkin paraconductivity is anticipated (on the conducting
side) and in fact
observed experimentally in PbBi/Ge films (Hsu, Chervenak and Valles 1995).

For the MoGe thin films, Yazdani and Kapitulnik (Yazdani and Kapitulnik 1995)
have also noted the surprising correlation between $H_c$ and
$H_{c2}$.  In light of this, and the similar trends in other
homogeneously disordered films, we conclude that the transition
at $H_c$ leads to a proliferation
of electronic excitations.  We interpret the insulator above $H_c$ as a
localized electron insulator rather than a Mott insulator of Cooper pairs.
We argue that it is at the lower field, $H_M$, that phase
coherence is lost and a Bose metal phase obtains as has
been discussed previously
(Dalidovich and Phillips 2000, Das and Doniach 1999). Consequently,
we are led to the global phase diagram shown in Fig. (\ref{pdiag})
for the destruction of superconductivity in homogeneously disordered
thin films.  The presence of
the Bose metal phase (which results directly from
finite temperature relaxation between Cooper pairs (Dalidovich and
Phillips 2000)) represents direct evidence that the phase and
amplitude degrees of freedom are destroyed at different energy scales.
In the case of pair breaking, gapless electronic excitations exist on
the insulating side of the transition.
Consequently, the universality class of the transition is fundamentally
different (Fradkin 2001) than in the phase-only model.  As the
universality class  is no longer the 3D XY model, a critical resistance
of $R_Q=h/4e^2$ and a correlation length exponent of $\nu=2/3$
(for the clean system) need no longer hold.  This
change in universality class could explain the lack of a unique value for
the critical resistance in homogeneously disordered thin films and the
scatter in the experimental values for the product of the dynamical
and correlation length exponents
(Liu and Goldman 1994, Seidler, Rosenbaum and Veal 1992,
Markovic, Christiansen and Goldman 1998). Progress in including
normal electrons into the IST could be made along the lines pursued
recently by Feigelman and Larkin
(Feigelman and Larkin 1998).  Finally, if our argument
is correct that phase coherence is lost at $H_M$, then the superfluid
density should exhibit the characteristic 
drop-off (Berezinskii 1971, Kosterlitz and Thouless 1973) to zero
at $H_M$.  Experiments which probe the penetration depth,
though extremely difficult on thin films,
are essential for these unanswered questions to be laid to rest.

\acknowledgements
This work was funded by the DMR
of the NSF grant No. DMR98-96134.  We would especially
like to acknowledge
the invaluable advice and encouragement of Ali Yazdani, R. Dynes and J.
Valles. J. Valles
also provided us with his unpublished data (Figs. 3 and 4). We would also
like to thank N. Markovic for the data points which led to Figure (2) and
Eduardo
Fradkin for clarifying discussions on the universality class of the
pair-breaking
transition.  We have also benefitted from conversations with A.
Kapitulnik, A. Goldman,
A. Bezryadin, M. Tinkham, A. Schakel, S. Kivelson, and A. Hebard.

\vspace{1cm}

\begin{center}
References
\end{center}

Barber, R. P., Merchant, L. M., La Porta, A., and Dynes, R. C., 1994,
Phys. Rev. B {\bf 49}, 3409.

Bardeen, J., Cooper, L. N., and Schrieffer, J. R., 1957,
Phys.  Rev. {\bf 106}, 162; {\bf 108}, 1175.

Berezinskii, V. L., 1971, Zh. Eksp. Teor. Fiz. {\bf 61}, 1144
[Sov. Phys. JETP {\bf 34}, 610].

Chervenak, J. A., 1998, Ph. D. Thesis Brown University.

Chervenak, J. A., and Valles, J. M., 1999, Phys. Rev. B {\bf 59}, 11209.

Chervenak, J. A., and Valles, J. M., 2000, Phys. Rev. B, R9245.

Dalidovich, D., and Phillips, P., 2000, ``Interaction-induced
Bose Metal,'' cond-mat/0005119.

Damle, K., and Sachdev, S., 1997, Phys. Rev. B {\bf 56}, 8714.

Das, D., and Doniach, S., 1999, Phys. Rev. B {\bf 60} 1261.

Dynes, R. C., Garno, J. P., and Rowell, J. M., 1978, Phys. Rev. Lett.
{\bf 40}, 479.

Dynes, R. C., et al., 1994,  "Ordering Disorder: Prospect and Retrospect in
Condensed Matter Physics" edited by V.  Srivastava, A. K. Bhatnager and D. G.
Naugle, AIP Conf Proc. No. 286 (AIP, New York) pp. 96-108.

Efros, A. L., 1976, J. Phys. C {\bf 9}, 2021.

Ephron, D., Yazdani, A., Kapitulnik, A., and Beasley, M. R., 1996,
Phys. Rev. Lett. {\bf 76}, 1529.

Feigelman, M. V., and Larkin, A. I., 1998, Chem. Phys. {\bf 235}, 107.

Fisher, M. P. A., 1990, Phys. Rev. Lett. {\bf 65}, 923.

Fisher, M. P. A., Grinstein, G., and Girvin, S. M., 1990, Phys. Rev.
Lett. {\bf 64}, 587.

Fradkin, E., 2001, private communication.

Geerligs, L. J., et. al., 1989, Phys. Rev. Lett. {\bf 63}, 326.

Goldman, A. M., and Markovic, N., 1998, {\it Physics Today} {\bf 51}
(11) 39-44.

Hebard, A. F., and Paalanen, M. A., 1990, Phys. Rev. Lett. {\bf 65}, 927.

Hsu, S. Y., 1995, Ph D. Thesis, Brown University.

Hsu, S. Y., Chervenak, J. A., and Valles, J. M., 1995,
Phys. Rev. Lett. {\bf 75}, 132.

Jaeger, H. M., et. al., 1989, Phys. Rev. B {\bf 40}, 182.

Kosterlitz, J. M., and Thouless, D. J., 1973, J. Phys. C {\bf 6}, 1181.

Kouh, T.,1999, unpublished.

Liu, Y., and Goldman, A. M., 1994, Mod. Phys. Lett. {\bf 8}, 277.

Markovic, N., Christiansen, C., and Goldman, A. M., 1998, Phys. Rev.
Lett. {\bf 81}, 5217.

Mason, N., and Kapitulnik, A.,1999, Phys. Rev. Lett. {\bf 82},
5341.

Mason, N., and Kapitulnik, A., 2000, cond-mat/0006138.

Merchant, L., Ostrick, J., Barber, R. P., and
Dynes, R. C., 2001, Phys. Rev. B {\bf 63}, 134508.

Schakel, A. M. J., 2000, Acta Phys. Pol. B 31, 2899.

Seidler, G. T., Rosenbaum, T. V., and Veal, B. W., 1992,
Phys. Rev. B {\bf 45}, 10162.

Valles, J. M., Dynes, R. C., and Garno, J. P., 1992, Phys.
Rev. Lett. {\bf 69}, 3567.

Valles, J. M., Chervenak, J. A., Hsu, S. Y.,
and Kouh, T. J., 2000, ``Fluctuation Effects in High Sheet Resistance
Superconducting Films,'' cond-mat/0010114.

Yazdani, A., and Kapitulnik, A., 1995, Phys. Rev. Lett.{\bf 74}, 3037.

\end{document}